\documentclass[a4paper,11pt]{article}
\pdfoutput=1 
\usepackage{jinstpub} 

\usepackage{lineno}

\title{\boldmath Long-Delayed Afterpulse Measurement of JUNO 20-inch Photomultiplier Tubes}

\author[a,e,*]{Xiaojie Luo\note[*]{Corresponding author.}}
\author[c]{, Cailian Jiang}
\author[a,b]{, Haojie Dong}
\author[a,\dagger]{, Yuduo Guan\note[\ensuremath{\dagger}]{Corresponding author.}}
\author[a]{, Gaosong Li}
\author[a]{, Zhonghua Qin}
\author[a,b]{, Zhenning Qu}
\author[a,b]{, Junyu Shao}
\author[a,b]{, Mingxia Sun}
\author[a]{, Liangjian Wen}
\author[a]{, Zeyuan Yu}
\author[d]{, Boyi Zheng}

\affiliation[a]{Institute of High Energy Physics, Chinese Academy of Sciences,\\
  19B Yuquan Road, Beijing 100049, China}
\affiliation[b]{School of Physical Sciences, University of Chinese Academy of Sciences,\\
  19A Yuquan Road, Beijing 100049, China}
\affiliation[c]{School of Physics, Nanjing University,\\
  22 Hankou Road, Nanjing 210093, China}
\affiliation[d]{University of Science and Technology of China,\\
  96 Jinzhai Road, Hefei 230026, China}
\affiliation[e]{China Center of Advanced Science and Technology,\\
  Beijing 100190, China}

\emailAdd{luoxj@ihep.ac.cn}
\emailAdd{guanyuduo@ihep.ac.cn}

\abstract{In large-scale liquid scintillator detectors such as the Jiangmen Underground Neutrino Observatory (JUNO), high-intensity events like cosmic muons induce photomultiplier tube (PMT) afterpulses that can interfere with the analysis of delayed physics signals. To systematically evaluate this instrumental background, we present a dedicated measurement of long-delayed afterpulses in two types of JUNO 20-inch PMTs: a dynode-based PMT and a microchannel-plate (MCP) PMT. The afterpulse time profiles were first characterized within a direct 1.8~ms waveform window and were further extended to 20~ms using a sliding-window readout strategy. Distinct long-delayed components are observed, revealing a strong dependence on the PMT multiplication structure. The dynode PMT exhibits a broad afterpulse component peaking at approximately 260~$\mu$s, whereas the MCP-PMT shows a pronounced peak around 90~$\mu$s, an additional component around 550~$\mu$s, and a much smaller, broadly distributed millisecond-scale component. In selected microsecond-scale windows, the fitted afterpulse yields are at the $10^{-3}$--$10^{-2}$ level per primary photoelectron, while the mean afterpulse count per trigger generally increases with primary light intensity. The accumulated delayed activity can therefore become non-negligible following high-intensity events. These quantitative findings provide inputs for PMT response characterization and for modeling delayed correlated backgrounds in high-precision neutrino experiments.}


\begin{document}
\maketitle
\flushbottom

\section{Introduction}
\label{sec:intro}

Photomultiplier tubes are sensitive single-photon detectors first developed in the 1930s \cite{kubetsky1937,zworykin1936}. Among the various PMT architectures, conventional dynode-based PMTs and microchannel-plate PMTs are widely used \cite{hamamatsu2006}. PMTs have found widespread application in particle physics \cite{RENKER2009207,tsoulfanidis2010} and medical imaging techniques \cite{anger1958,knoll1999}. Large-scale scintillation and Cherenkov detectors use PMTs to measure low-energy neutrinos \cite{Alimonti:2009aa,Eguchi:2002nm,An:2015kca} and extremely high-energy astrophysical events \cite{IceCube:2013cd,LHAASO:2021gok,KM3NeT:2025aa}. These precision experiments require a thorough understanding of the PMT response.

An important instrumental response in PMTs is afterpulsing, namely the occurrence of secondary pulses following the primary photoelectron signal \cite{PhysRev.84.1248}. Afterpulses are commonly associated with ion feedback due to residual-gas ionization \cite{PBCoates_1973,P_B_Coates_1973,Akchurin:2007ion_afterpulse}, while delayed electron emission from PMT surfaces may produce additional delayed components \cite{Morozov:2023delayed_emission}. In large liquid-scintillator detectors, high-energy events such as cosmic muons generate intense PMT signals; the resulting afterpulses form a correlated instrumental component in the post-muon time region. Dedicated measurements are therefore required to assess their impact on event reconstruction, calibration, and muon veto strategies.

The time scales relevant to this study extend beyond the conventional few-microsecond afterpulse region. Spallation neutrons produced by cosmic muons provide an important calibration sample for neutron-related detector response.   Neutron capture on hydrogen releases a $\sim$2.2 MeV $\gamma$ ray with a characteristic capture time of about 220 $\mu$s. Long-delayed PMT afterpulses in the same post-muon time region may add instrumental charge to neutron-capture events and thereby distort the reconstructed energy spectrum.

The afterpulse time distribution also provides input for defining post-muon veto windows. Large liquid-scintillator neutrino detectors commonly use millisecond-scale detector-wide or localized muon veto strategies to control muon-induced correlated backgrounds \cite{Alimonti:2009aa,Eguchi:2002nm,An:2015kca,juno_ppnp,JUNO:2025first_osc,Genster_2020}. If the veto duration is insufficient, residual afterpulses from large PMT signals can survive into subsequent physics windows and bias energy reconstruction. Millisecond-scale afterpulse measurements are therefore needed to constrain this instrumental contribution.

Many studies have investigated afterpulses in PMTs, with the majority focusing on short-delayed components occurring within a few microseconds and amounting to a few percent of the primary signal \cite{Zhao_2016}. Additionally, several investigations have identified longer-delayed afterpulses on the order of hundreds of microseconds, with broad delayed-time distributions \cite{POLESHCHUK2012362,TUDYKA201639,IceCube:2025paj}. 

JUNO is a large liquid-scintillator neutrino detector equipped with 17,612 20-inch PMTs and 25,600 3-inch PMTs \cite{juno_initial_performance_2026,juno_ppnp,An:2015kca}. Its high light yield and photocathode coverage enable precision reactor-neutrino measurements, including neutrino mass ordering and oscillation-parameter studies \cite{juno_reactor_nmo,JUNO:2025first_osc}, and provide sensitivity to solar, atmospheric, supernova, and geoneutrinos \cite{juno_ppnp}. Precise afterpulse measurements of JUNO 20-inch PMTs, namely the HPK R12860 and NNVT GDB-6201, are therefore needed to understand the after-muon detector response. Many related afterpulse measurements have been reported; however, few have investigated sub-percent components at delays beyond 20~$\mu$s \cite{Zhao:2022gks,WU2021165351,Liu:2025evh,abusleme_mass_2022}.

In this work, we measure long-delayed afterpulsing at the sub-percent level in two 20-inch PMT models used in the JUNO experiment, employing millisecond-scale waveform readout. The measurement targets afterpulses occurring more than 10~$\mu$s after the primary pulse and extends the accessible delay range up to 20~ms.

Sections~\ref{sec:hardware} and~\ref{sec:data_processing} describe the experimental setup and data analysis, respectively. Section~\ref{sec:results} presents the afterpulse time profiles and intensity dependence, followed by a discussion of systematic effects and possible mechanisms in Section~\ref{sec:discuss}.


\section{Experimental Setup}
\label{sec:hardware}

\subsection{Optical Setup and Data Acquisition}
\label{sec:experimental_configuration}

\begin{figure}
\centering
\includegraphics[width=1\linewidth]{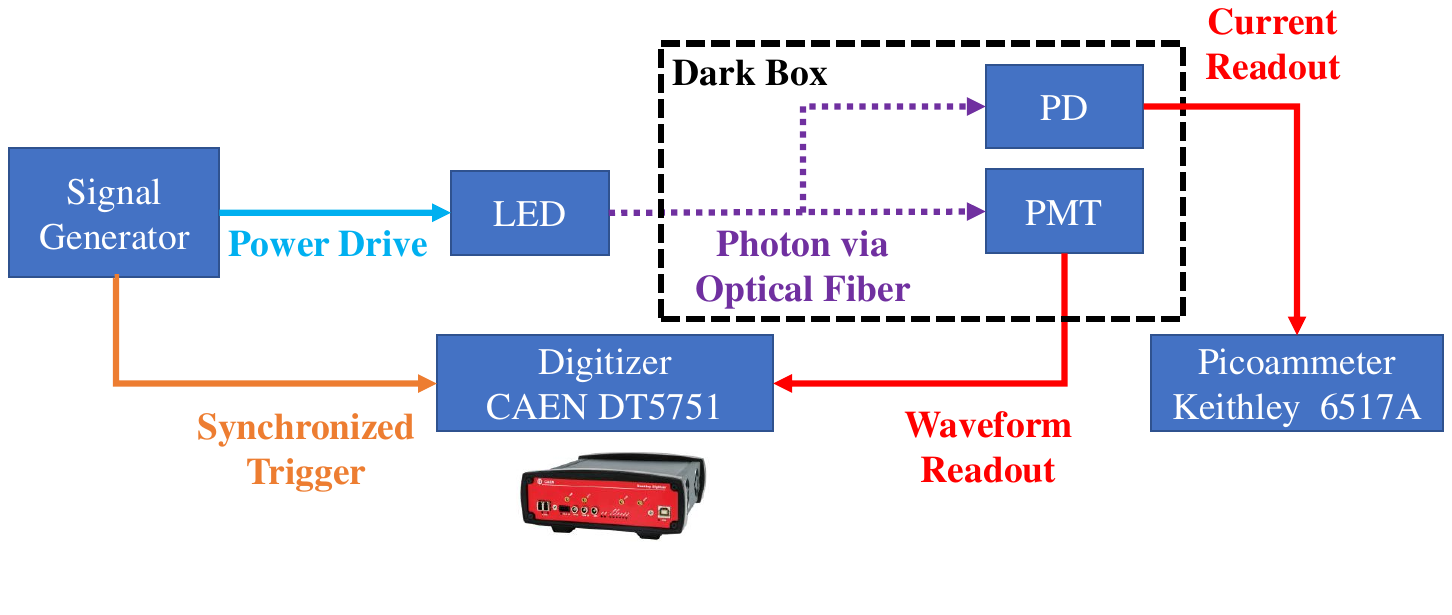}
\caption{Schematic of the experimental setup. A fiber splitter directs light to the PMT and to a Hamamatsu S1337-1010BR photodiode monitored with a Keithley 6517A picoammeter. The signal generator also triggers the digitizer.}
\label{fig:Experimental Layout}
\end{figure}

\begin{figure}
\centering
\includegraphics[width=0.5\linewidth]{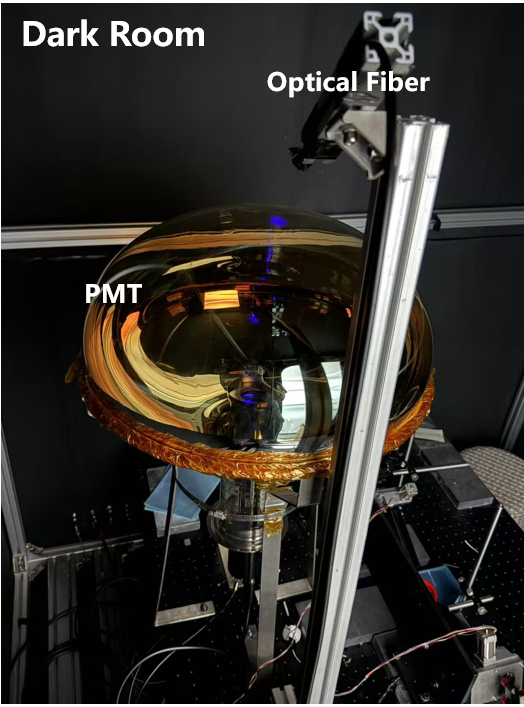}
  \caption{
  Photograph of the practical experimental setup for the afterpulse measurement.
  The 20-inch PMT was placed inside a light-tight dark box and illuminated by the 430~nm
  fiber-coupled LED through an optical fiber. The main components, including the
  PMT and the optical fiber, are indicated in the photograph.
  }
\label{fig:Pratical Layout}
\end{figure}

To measure long-delayed afterpulses with low occurrence probability, light inputs ranging from a few hundred to several thousand photoelectrons (PE) were applied to the PMT. The setup is shown in Figure~\ref{fig:Experimental Layout}. A signal generator drove a 430~nm fiber-coupled LED (Thorlabs M430F1). The LED pulse width was 1~$\mu$s for the direct 1.8~ms measurements and 5~$\mu$s for the 20~ms sliding-window measurements. Measurements were performed at $23\pm1\,^{\circ}$C. A fiber splitter directed one branch to the PMT and the other to a Hamamatsu S1337-1010BR photodiode (PD). The PD current was read by a Keithley 6517A picoammeter and corrected by subtracting an LED-off measurement acquired before each scan. The illumination spot on the photocathode was approximately 1~cm in diameter. PMT waveforms were digitized at 1~GS/s with a CAEN DT5751 over a 1.8~ms window. A photograph of the setup is shown in Figure~\ref{fig:Pratical Layout}.

LED-on and LED-off measurements were compared to evaluate external background and PMT dark noise. Intensity scans measured the dependence of the afterpulse yield on the primary signal.


We characterized the two PMTs listed in Table~\ref{tab:pmt_specification} through SPE calibration, dark-noise counting, and afterpulse measurements. The gain values listed in the table are derived from the single-photoelectron charge calibration described in Sec.~\ref{sec:data_processing}.

Both PMTs were operated under positive high voltage using passive resistor-chain voltage dividers based on the JUNO BX2 scheme; no active voltage-regulation or feedback components were used. Detailed schematics of the corresponding MCP-PMT and dynode-PMT divider circuits are shown in Figure 20 of Ref. \cite{Luo_2025}.

\begin{table}[htbp]
    \centering
    \caption{Summary of the 20-inch PMTs used for afterpulse measurements.}
    \label{tab:pmt_specification}
    \begin{tabular}{lccc}
        \hline
        PMT type & Manufacturer and model & Operating voltage & Peak Gain \\
        \hline
        Dynode PMT
        & HPK R12860
        & 1800 V
        & $(1.071 \pm 0.004) \times 10^{7}$
        \\
        MCP-PMT
        & NNVT GDB-6201
        & 1750 V
        & $(8.95 \pm 0.13) \times 10^{6}$
        \\
        \hline
    \end{tabular}
\end{table}


\subsection{Sliding-Window Readout Strategy}
\label{sec:window_extension}

\begin{figure}[htbp]
  \centering
  \includegraphics[width=1\textwidth]{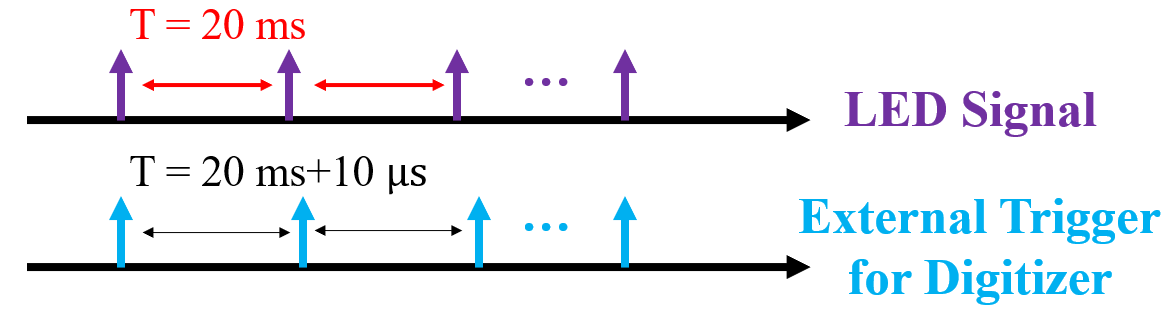}
\caption{
Principle of the sliding-window strategy used to extend the effective afterpulse
measurement window. The LED emission period is set to 20~ms and the digitizer
trigger period to 20.01~ms, causing the relative phase between the
LED signal and the digitizer trigger to drift over successive measurements.
A short 10~$\mu$s waveform window is used for each trigger, and the acquisition
windows progressively sample the designed 20~ms afterpulse time range after
accumulating many LED flashes.
}
\label{fig:sliding_window_strategy}
\end{figure}

With the optical setup and data-acquisition configuration described above, the maximum waveform
recording window available from the digitizer is limited to 1.8~ms for each
trigger. To extend the measurement to longer time scales, a sliding-window
strategy was implemented, as illustrated in Figure~\ref{fig:sliding_window_strategy}. Since this strategy relies on stable and continuous
data taking over many LED flashes, a short waveform recording window of
10~$\mu$s was used for each trigger to minimize possible trigger losses during
the scan.

In this strategy, the LED emission period is set to 20~ms, whereas the
external trigger period for the digitizer is set to 20.01~ms.
As a result, the relative phase between the LED signal and the digitizer trigger
gradually changes over successive LED
flashes, allowing the short acquisition window to sample different delay times
after the primary light pulse. By accumulating many measurements, the
10~$\mu$s waveform windows progressively cover the designed 20~ms afterpulse
time range. The full 20~ms delay axis was reconstructed according to the
trigger sequence, with each waveform assigned to its corresponding phase offset
within the 20~ms cycle. The event time stamps were used to check the continuity
of the trigger stream and to identify possible missing triggers.

\section{Waveform Reconstruction and Data Analysis}
\label{sec:data_processing}

\begin{figure}[htbp]
  \centering
  \includegraphics[width=1\textwidth]{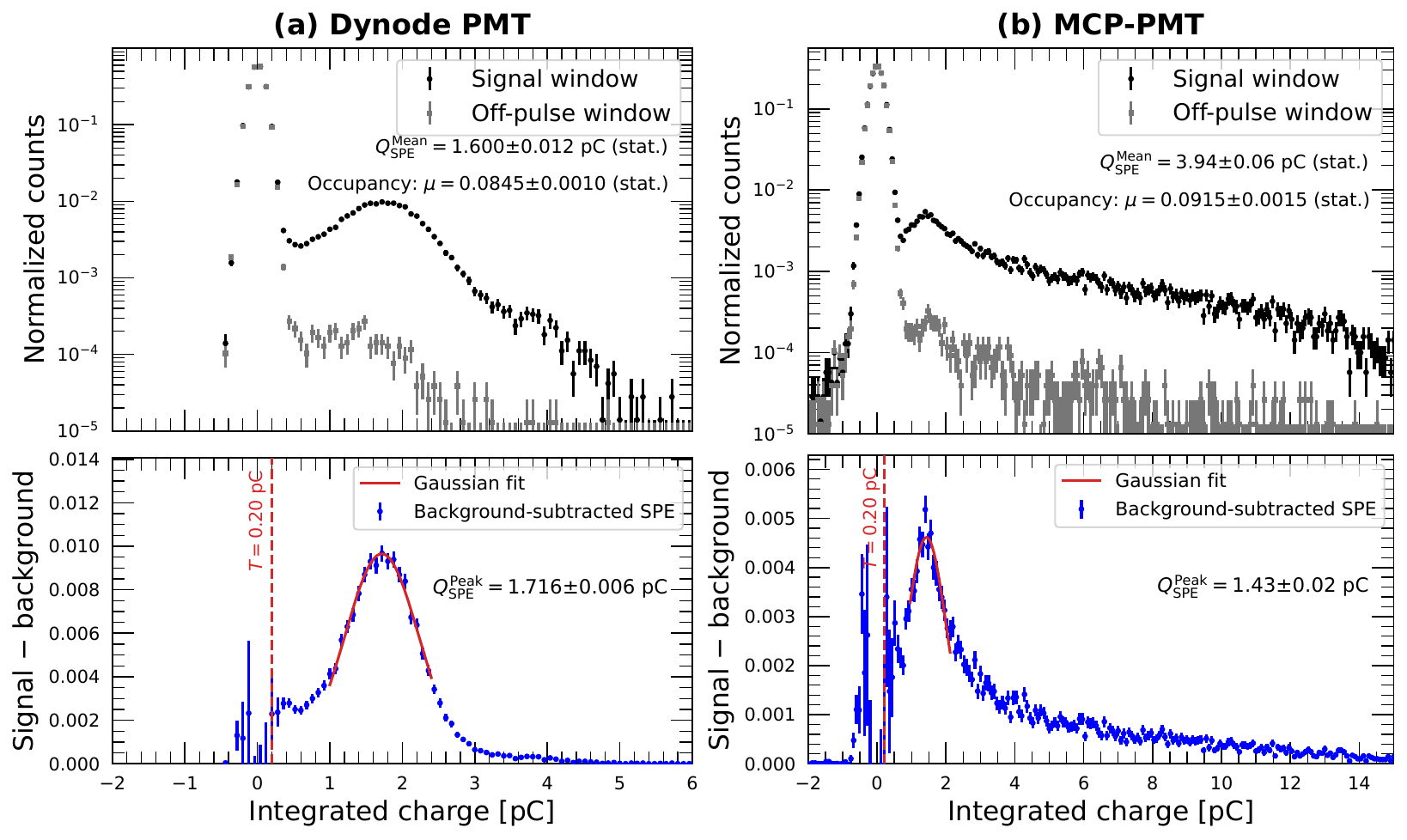}
  \caption{SPE charge spectra for (a) the dynode PMT and (b) the MCP-PMT. The upper panels compare the signal-window and same-trigger off-pulse-window spectra, normalized by their respective negative-charge counts, and give the mean SPE charge and occupancy with paired-bootstrap statistical errors. The lower panels show the difference and the Gaussian fit to the SPE peak; red dashed lines mark the occupancy threshold. The quoted peak-charge uncertainties are statistical fit errors.}
  \label{fig:spe_peak_calibration}
\end{figure}

A single-photoelectron (SPE) calibration was performed with low-intensity pulses from a 405~nm laser (PicoQuant LDH-D-C-405), with a pulse width below 50~ps, permitting a narrow charge-integration window. Signal-window charge spectra were compared with equal-duration off-pulse windows from the same acquisition. Each spectrum was normalized by its negative-charge pedestal count ($Q<0$) before subtraction. A Gaussian was fitted to the resulting SPE peak, with its mean denoted by $Q_{\mathrm{SPE}}^{\mathrm{Peak}}$. The spectra in Figure~\ref{fig:spe_peak_calibration} give peak charges of $1.716\pm0.006$~pC for the dynode PMT and $1.434\pm0.021$~pC for the MCP-PMT. The corresponding Peak Gain values in Table~\ref{tab:pmt_specification} are calculated as $Q_{\mathrm{SPE}}^{\mathrm{Peak}}/e$.

Following the model-independent calibration method of Saldanha et al.~\cite{Saldanha:2017spe}, the mean SPE charge used to calibrate the primary NPE is estimated as
\begin{equation}
  Q_{\mathrm{SPE}}^{\mathrm{Mean}}(T)=
  \frac{\langle q\rangle_{\mathrm{sig}}-\langle q\rangle_{\mathrm{off}}}{\widehat\mu(T)},
  \label{eq:mean_spe_calibration}
\end{equation}
where $\widehat\mu(T)$ is the occupancy (the mean number of photoelectrons per trigger), estimated as
\begin{equation}
  \widehat\mu(T)=\ln\frac{f_{\mathrm{off}}(T)}{f_{\mathrm{sig}}(T)}.
  \label{eq:spe_occupancy}
\end{equation}
Here $q$ is the pedestal-centered integrated charge. The subscripts $\mathrm{sig}$ and $\mathrm{off}$ denote the signal window and the equal-duration off-pulse window from the same trigger, respectively; their pedestals are centered separately. We adopt a fixed occupancy threshold $T=0.20$~pC for both PMTs throughout this analysis, marked in Figure~\ref{fig:spe_peak_calibration}; $f_{\mathrm{sig}}(T)$ and $f_{\mathrm{off}}(T)$ are the fractions with $q<T$, each normalized by its accepted trigger count. All triggers passing the baseline-quality selection enter the charge means without a charge cut.
This gives $Q_{\mathrm{SPE}}^{\mathrm{Mean}}=1.600\pm0.012$~pC for the dynode PMT and $3.94\pm0.06$~pC for the MCP-PMT. Statistical errors are the standard deviations of 2,000 paired event-bootstrap replicas that recompute the pedestal centers, charge means and occupancy. The resulting mean-SPE statistical errors are propagated to all primary NPE values. Underamplified photoelectrons below threshold can bias the occupancy low and the mean SPE charge high~\cite{Saldanha:2017spe}.

The measured primary NPE, including that in the 20~ms measurement, is obtained from the integrated primary-pulse charge divided by $Q_{\mathrm{SPE}}^{\mathrm{Mean}}$. Its statistical uncertainty combines the primary-charge mean error and the paired-bootstrap SPE calibration error in quadrature. The latter contributes relative uncertainties of 0.74\% for the dynode PMT and 1.60\% for the MCP-PMT.

\begin{figure}[htbp]
  \centering
  \includegraphics[width=1\textwidth]{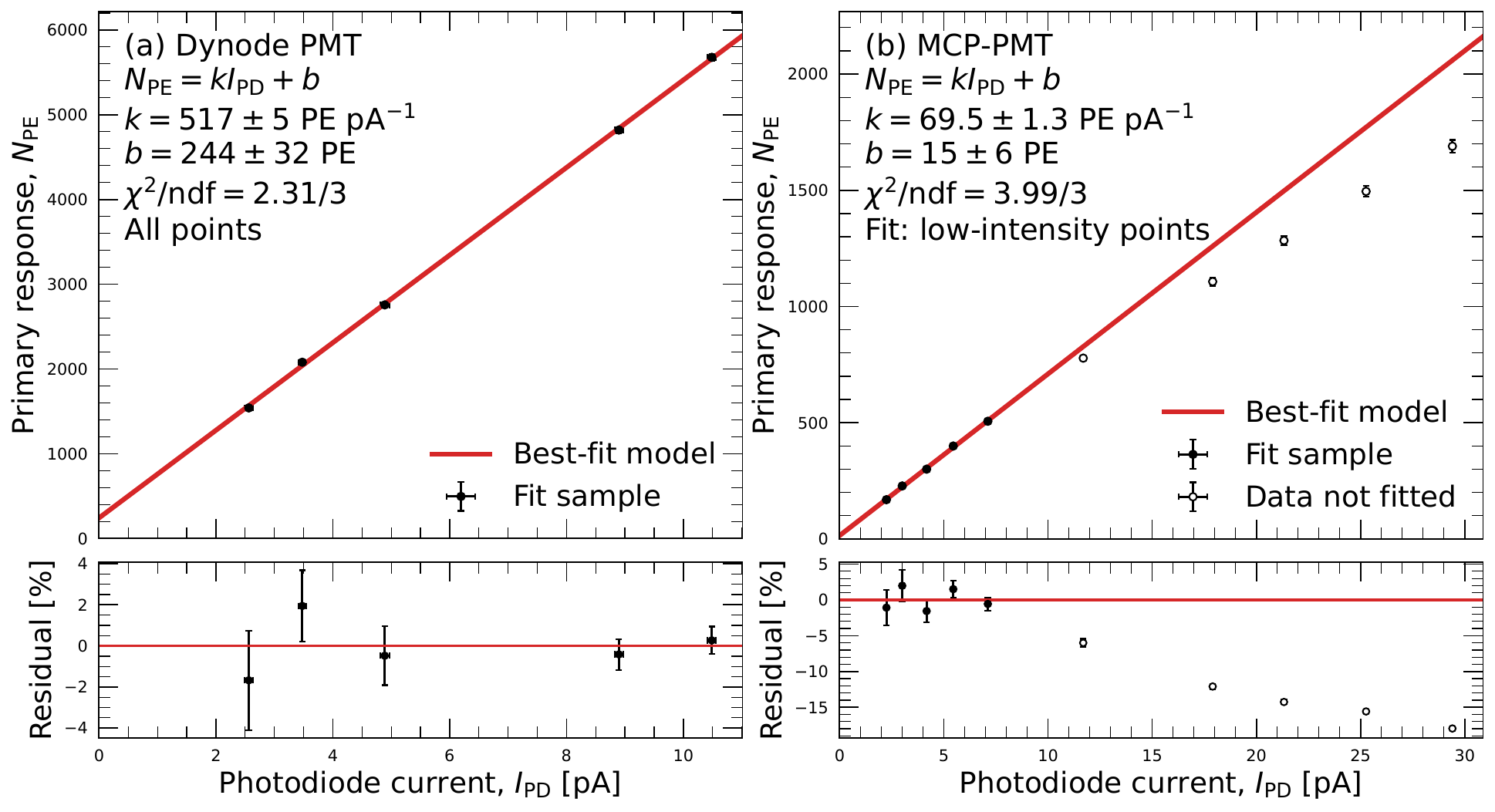}
  \caption{Primary NPE versus dark-current-subtracted photodiode current, used to characterize PMT nonlinearity. Filled points enter the affine low-intensity fits and open points are shown for comparison. The dynode-PMT response is linear over the measured range. For the MCP-PMT, the low-intensity relation is used to correct primary NPE of the afterpulse data. Horizontal bars show PD-current statistics; vertical bars combine primary-charge mean statistics and the SPE calibration statistics. Residual bars propagate pointwise charge and PD-current statistics; the common SPE scale cancels in the relative residual.}
  \label{fig:pd_calibration}
\end{figure}

For larger primary signals, we also check whether PMT nonlinearity affects this charge calibration. Figure~\ref{fig:pd_calibration} characterizes PMT nonlinearity by comparing primary NPE with the dark-current-subtracted PD current, which monitors the incident light intensity. Extrapolation of the low-intensity linear relation to higher PD currents provides a reference for evaluating the nonlinear response. The dynode PMT is linear over the measured range. For the MCP-PMT, the nonlinearity coefficient is $C_{\mathrm{NL}}=N_{\mathrm{PE}}^{\mathrm{meas}}/(kI_{\mathrm{PD}}+b)=1+r/100$, where $kI_{\mathrm{PD}}+b$ is the low-intensity linear reference and $r$ is the percent residual in the lower panel of Figure~\ref{fig:pd_calibration}. For each direct-window measurement, we obtain the coefficient by interpolating the calibration curve at the measured primary NPE and apply $N_{\mathrm{PE}}^{\mathrm{corr}}=N_{\mathrm{PE}}^{\mathrm{meas}}/C_{\mathrm{NL}}$. Dynode results use the directly measured primary NPE.

\begin{figure}[htbp]
  \centering
  \includegraphics[width=1\textwidth]{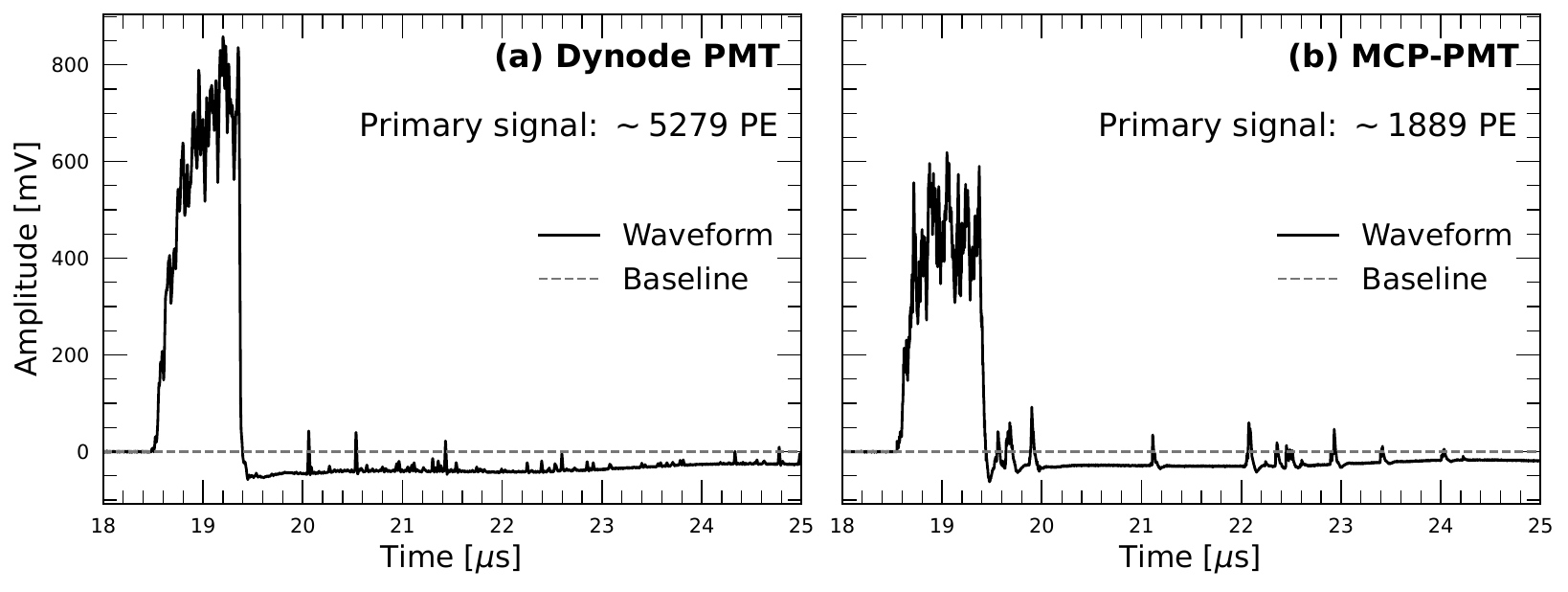}
\caption{
Representative raw waveforms for (a) the HPK R12860 dynode PMT and (b) the NNVT GDB-6201 MCP-PMT. The primary NPE values are approximately 5279 PE and 1889 PE, respectively; the latter uses the PD-based linearity correction. Dashed lines indicate the waveform baseline.
}
\label{fig:raw_waveform_display}
\end{figure}

\begin{figure}[htbp]
\centering
\includegraphics[width=1\textwidth]{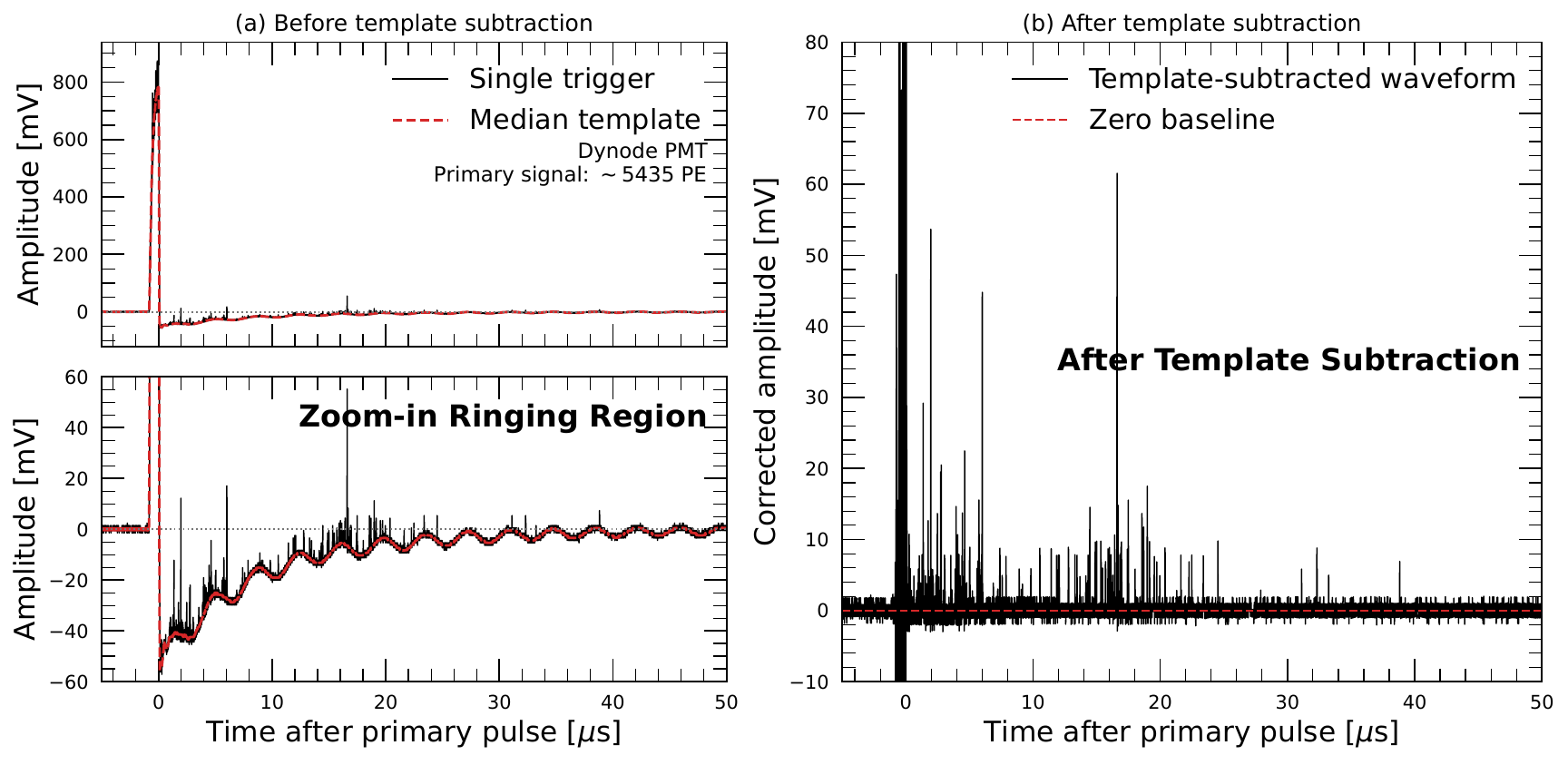}
\caption{Example of baseline correction for a dynode-PMT trigger at approximately $5.4\times10^3$ primary NPE. The left panels show the single-trigger waveform (black) and the pointwise median template formed from a subset of triggers (red dashed), in full and enlarged amplitude ranges. The right panel shows their difference before pulse finding; the dashed red line marks the zero baseline. The primary-pulse peak defines $t=0$.}
\label{fig:waveform_reconstruction}
\end{figure}

Representative waveforms near the upper end of the analyzed range are shown in Figure~\ref{fig:raw_waveform_display}. The primary sizes are approximately $5.28\times10^3$ PE for the dynode PMT and $1.89\times10^3$ PE for the MCP-PMT. At each light setting, a template formed from the pointwise median of a subset of triggers was subtracted before pulse finding, as demonstrated in Figure~\ref{fig:waveform_reconstruction}. Pulses were identified by peak amplitude using fixed thresholds equivalent to $0.708\pm0.006$ peak PE for the dynode PMT and $0.582\pm0.011$ peak PE for the MCP-PMT. For each accepted pulse, its integrated charge and time relative to the primary-pulse peak were recorded.

\begin{figure}
    \centering
    \includegraphics[width=1\linewidth]{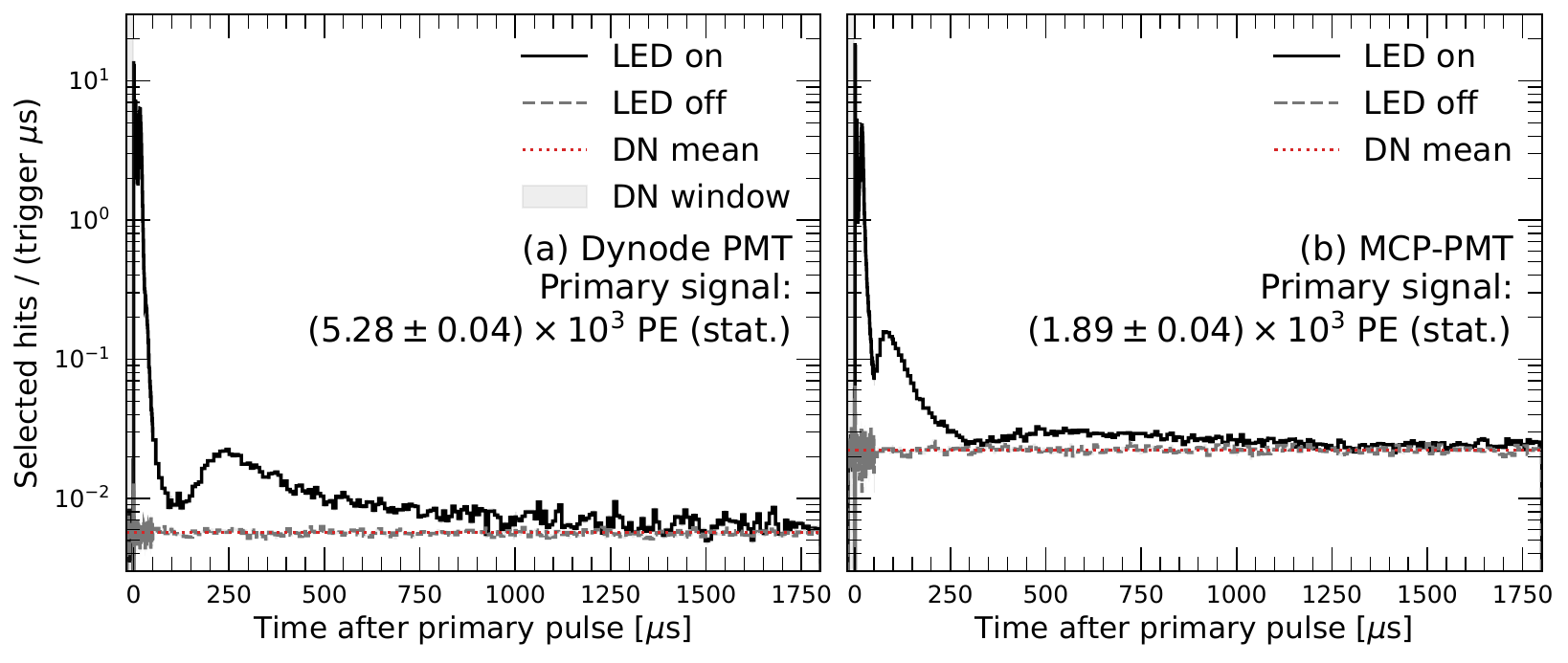}
    \caption{
    Reconstructed timing distributions for (a) the dynode PMT and (b) the MCP-PMT before dark-noise subtraction. The quoted primary NPE values give the calibrated mean and its statistical uncertainty, including the SPE calibration statistics. The MCP-PMT value also includes PD-mapping statistics. The profile bands show hit-count statistics.
    }
    \label{fig:afterpulse_raw}
\end{figure}

The primary timing peak was aligned to $t=0$~ns. The reconstructed hit times were then histogrammed and normalized by the accepted trigger count and bin width. Figure~\ref{fig:afterpulse_raw} shows the distributions before dark-noise subtraction. Both PMTs exhibit a clear delayed excess above the scaled LED-off spectrum. The absence of corresponding structures in the LED-off controls disfavors an origin in uncorrelated dark noise.

\section{Results}
\label{sec:results}

\subsection{Afterpulse Time Profiles up to 1.8 ms}
\label{sec:afterpulse_profile_direct}

\begin{figure}[htbp]
\centering
\includegraphics[width=1\linewidth]{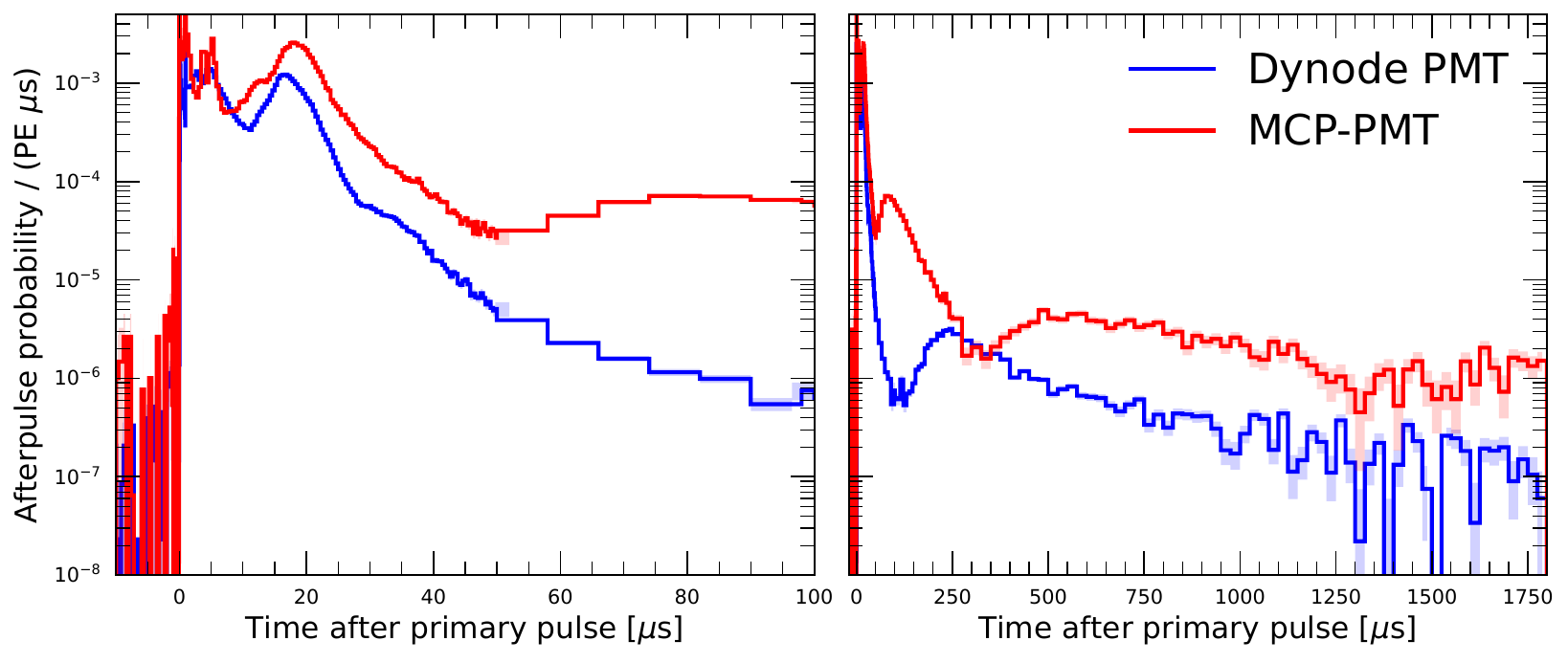}
\caption{
Afterpulse time profiles for the dynode PMT and MCP-PMT within the direct 1.8~ms waveform window. Dark noise estimated from the pre-primary control interval is subtracted eventwise. The left and right panels show the early and full ranges, respectively. The MCP-PMT primary NPE is corrected with the PD response in Figure~\ref{fig:pd_calibration}. The distributions are normalized by primary NPE and bin width. Bands show profile and dark-noise statistics.
}
\label{fig:Results of After Pulse}
\end{figure}

Following the data processing described in Sec.~\ref{sec:data_processing},
the afterpulse time profiles were constructed for large primary light signals
within the 1.8~ms DAQ window. With this time origin, the residual background level was estimated from
the $[-18,-2]~\mu$s pre-primary region of each LED-on sample and subtracted. The resulting afterpulse
profiles for the dynode PMT and the MCP-PMT are shown in
Figure~\ref{fig:Results of After Pulse}. The distributions are normalized by the
calibrated number of photoelectrons in the primary pulse and by the bin width,
so that the vertical scale represents the afterpulse probability density per
primary photoelectron.

In the early-delay region, both PMTs show afterpulse components within the
first 100~$\mu$s after the primary pulse. Similar short-delay components have been
reported in previous measurements of JUNO 20-inch PMTs ~\cite{Zhao:2022gks}.
In addition to these known components, this measurement reveals additional
long-delayed afterpulse components at later times. The dynode PMT shows a broad
component peaking at approximately 260~$\mu$s, followed by a slowly decreasing
tail. The MCP-PMT exhibits a more concentrated component around 90~$\mu$s and a
weaker structure near 550~$\mu$s. The different time profiles indicate that the
long-delayed response depends on the PMT multiplication structure, while both
PMT types can generate correlated afterpulses far beyond the conventional
short-afterpulse time region.

These observations indicate that, besides the previously studied afterpulse
components within several tens of microseconds, the 20-inch PMTs under study can also
produce long-delayed afterpulse components at the hundreds-of-microseconds to
millisecond time scale. Under the strong illumination used in this measurement,
the region very close to the primary pulse contains a high density of afterpulse
hits, where pulse pile-up and overlap can affect the pulse counting. Therefore,
the quantitative pulse-count analysis in this work starts from 10~$\mu$s and is
performed in the selected time windows shown below.

\subsection{Yield Dependence on Primary Signal Intensity}
\label{sec:light_intensity_dependence}

Based on the afterpulse components identified in the 1.8~ms DAQ window, the
yield dependence on the primary signal intensity was studied using
measurements with different LED intensities. As the calibrated primary signal
increases, the afterpulse components in the selected time windows become more pronounced,
indicating that these afterpulses are correlated with the primary light
signal rather than being residual dark-noise fluctuations.

For a quantitative comparison, the reconstructed afterpulse entries after
eventwise subtraction of the dark-noise count measured in the pre-primary control interval were integrated within selected time windows and
normalized by the accepted trigger count. The dark count was scaled by the ratio of the signal- and control-window widths. The statistical vertical uncertainty is the standard error of the resulting eventwise corrected count. This pulse-count yield measures the mean number of reconstructed afterpulse
entries per accepted trigger. Possible large-signal effects on the absolute yield are discussed
in Sec.~\ref{sec:large_signal_gain_recovery}.

\begin{figure}[htbp]
\centering

\makebox[\textwidth][c]{%
\begin{minipage}[t]{0.31\textwidth}
    \centering
    \includegraphics[width=\linewidth]{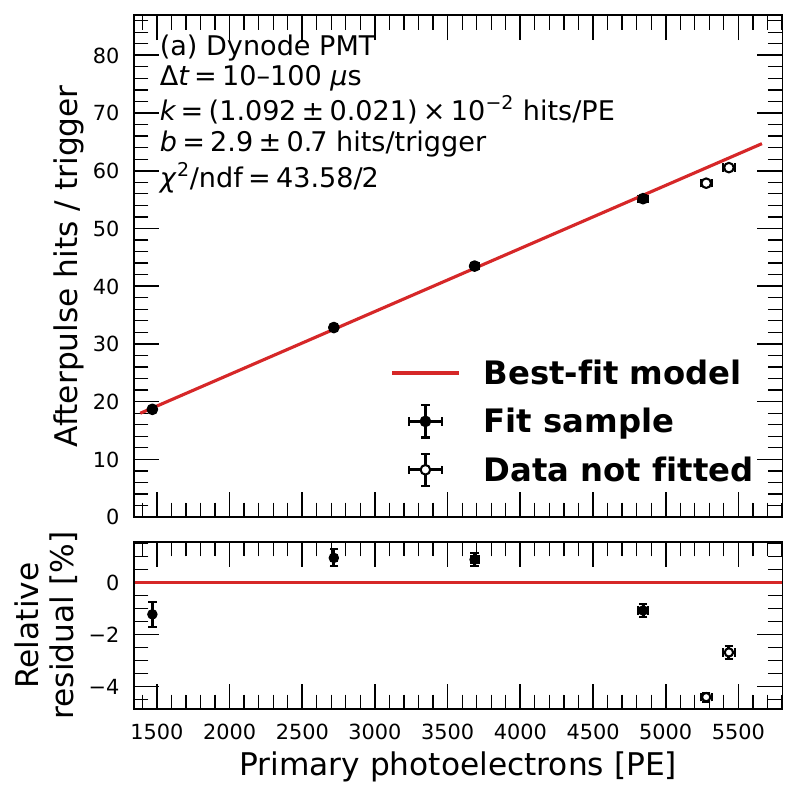}
    \\[1mm]
    (a) Dynode, 10--100~$\mu$s.
\end{minipage}
\hspace{0.06\textwidth}
\begin{minipage}[t]{0.31\textwidth}
    \centering
    \includegraphics[width=\linewidth]{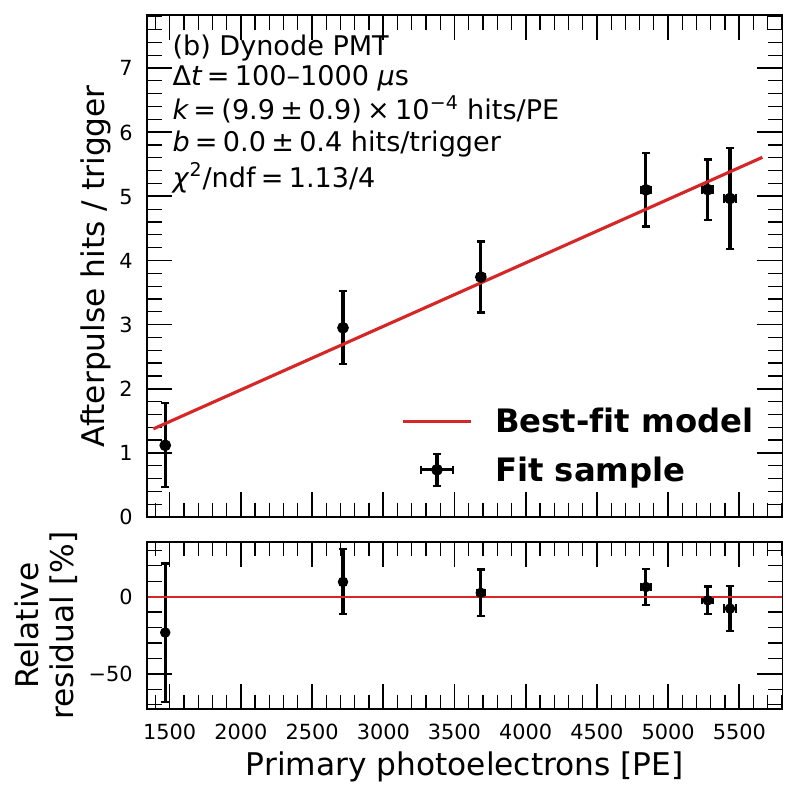}
    \\[1mm]
    (b) Dynode, 100--1000~$\mu$s.
\end{minipage}
}

\vspace{3mm}

\begin{minipage}[t]{0.31\textwidth}
    \centering
    \includegraphics[width=\linewidth]{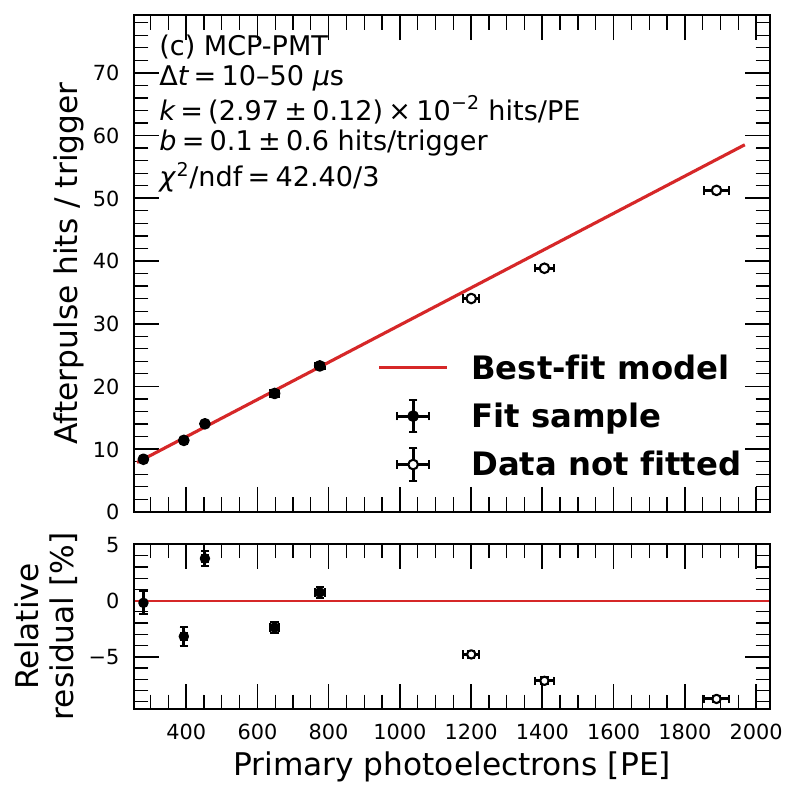}
    \\[1mm]
    (c) MCP, 10--50~$\mu$s.
\end{minipage}
\hfill
\begin{minipage}[t]{0.31\textwidth}
    \centering
    \includegraphics[width=\linewidth]{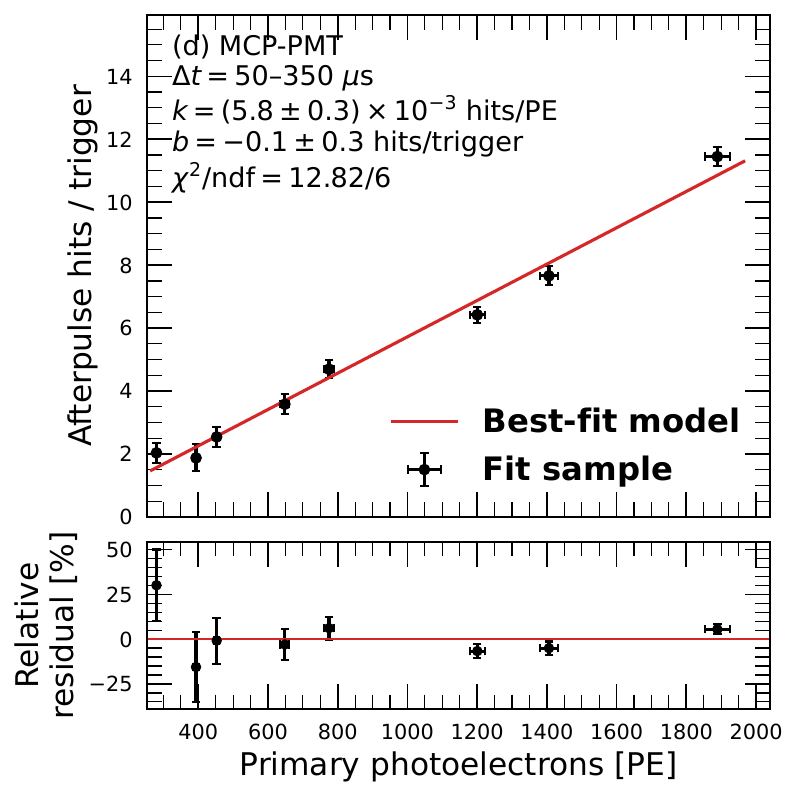}
    \\[1mm]
    (d) MCP, 50--350~$\mu$s.
\end{minipage}
\hfill
\begin{minipage}[t]{0.31\textwidth}
    \centering
    \includegraphics[width=\linewidth]{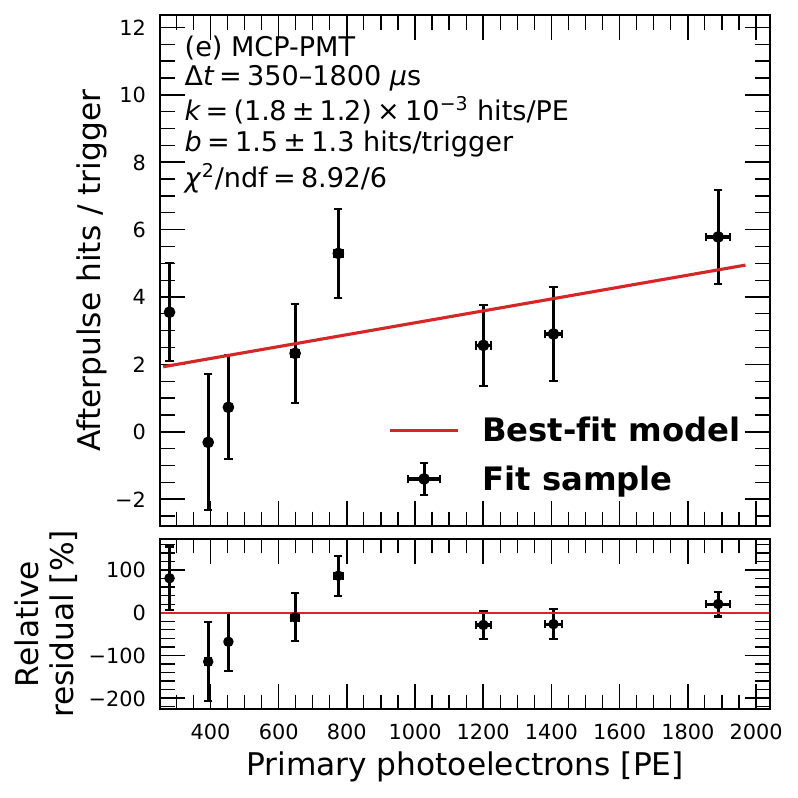}
    \\[1mm]
    (e) MCP, 350--1800~$\mu$s.
\end{minipage}

\caption{
Dependence of the afterpulse yield on the primary light intensity.
The afterpulse yield is defined as the number of reconstructed afterpulse
entries per accepted trigger after eventwise dark-noise subtraction. Red lines are affine ODR fits and lower panels show relative residuals, $100(y-y_{\mathrm{fit}})/y_{\mathrm{fit}}$, in percent. Filled points enter a fit and open points are outside the early-window fit selections. Vertical bars are statistical standard errors. Horizontal bars combine pointwise primary-NPE statistics with shared SPE calibration statistics; MCP-PMT bars also include shared PD dark-current and response-fit statistics. Only pointwise statistical uncertainties enter as independent ODR errors. The SPE uncertainty is a common scale across all primary NPE values; for the MCP-PMT, 10,000 fixed-seed pseudoexperiments jointly propagate the shared calibration contributions.
}
\label{fig:late_ap_linearity}
\end{figure}

For the dynode PMT, the 10--100~$\mu$s window covers the early afterpulse
component, while the 100--1000~$\mu$s window covers the broad long-delayed
component. For the MCP-PMT, three windows are considered: 10--50~$\mu$s for
the early afterpulse component, 50--350~$\mu$s for the pronounced component around
90~$\mu$s, and 350--1800~$\mu$s for the later delayed tail. The resulting
relationships between the afterpulse yield and the calibrated primary
photoelectrons are shown in Figure~\ref{fig:late_ap_linearity}. In these
windows, the afterpulse yield generally increases with primary light intensity. We fit $y=kx+b$ using orthogonal distance regression (ODR) with pointwise statistical uncertainties in both axes; shared calibration contributions are treated as described in the caption of Figure~\ref{fig:late_ap_linearity}.

The early-window fits use the four lowest-intensity dynode points and the five lowest-intensity MCP points to define a linear reference for extrapolation to higher intensities and assessment of nonlinearity; all measured points enter the other three fits. Even within these selected ranges, the dynode 10--100~$\mu$s and MCP 10--50~$\mu$s components deviate from a single affine response, with $\chi^2/\mathrm{ndf}=43.58/2$ and $42.40/3$, respectively. Large-signal voltage-divider and space-charge effects, and MCP channel-charge depletion followed by recharge, could change the effective gain and hence the number of pulses crossing threshold~\cite{hamamatsu2006,Wu:2023linearity}. The origin of this nonlinearity requires further study.

The fitted slopes for these time windows are summarized in
Table~\ref{tab:late_ap_yield}. The slopes
represent the number of afterpulse entries per trigger per primary
photoelectron. The measured values are at the $10^{-3}$--$10^{-2}$ level per primary
photoelectron. For the early-delay windows, the fitted slopes are empirical, fit-range-dependent descriptors. Therefore, although
the delayed contribution is small compared
with the prompt signal on a per-photoelectron basis, the accumulated delayed
activity can become non-negligible after high-intensity events containing many
photoelectrons.

\begin{table}[htbp]
\centering
\caption{
Afterpulse yields in selected time windows, defined as the slope $k$ of the affine relation between afterpulse entries per trigger and primary NPE. Uncertainties are statistical errors from the ODR fits.
}
\label{tab:late_ap_yield}
\smallskip
\begin{tabular}{ccc}
\hline
\textbf{PMT type} & \textbf{Time window} & \textbf{Fitted yield per primary PE} \\
\hline
Dynode PMT & 10--100~$\mu$s & $(1.092 \pm 0.021)\times 10^{-2}$ \\
Dynode PMT & 100--1000~$\mu$s & $(9.9 \pm 0.9)\times 10^{-4}$ \\
MCP-PMT   & 10--50~$\mu$s    & $(2.97 \pm 0.12)\times 10^{-2}$ \\
MCP-PMT   & 50--350~$\mu$s   & $(5.8 \pm 0.3)\times 10^{-3}$ \\
MCP-PMT   & 350--1800~$\mu$s & $(1.8 \pm 1.2)\times 10^{-3}$ \\
\hline
\end{tabular}
\end{table}

\subsection{Extended Time Profiles up to 20 ms}
\label{sec:extended_20ms_result}

To further examine afterpulse components beyond the 1.8~ms DAQ
window, the sliding-window method described in Sec.~\ref{sec:window_extension}
was used to extend the effective measurement range to 20~ms. In this method,
the relative phase between the LED signal and the digitizer trigger gradually
changes over successive measurements, so that the short 10~$\mu$s waveform
windows progressively sample the designed 20~ms delay range.

\begin{figure}[htbp]
\centering
\includegraphics[width=.75\textwidth]{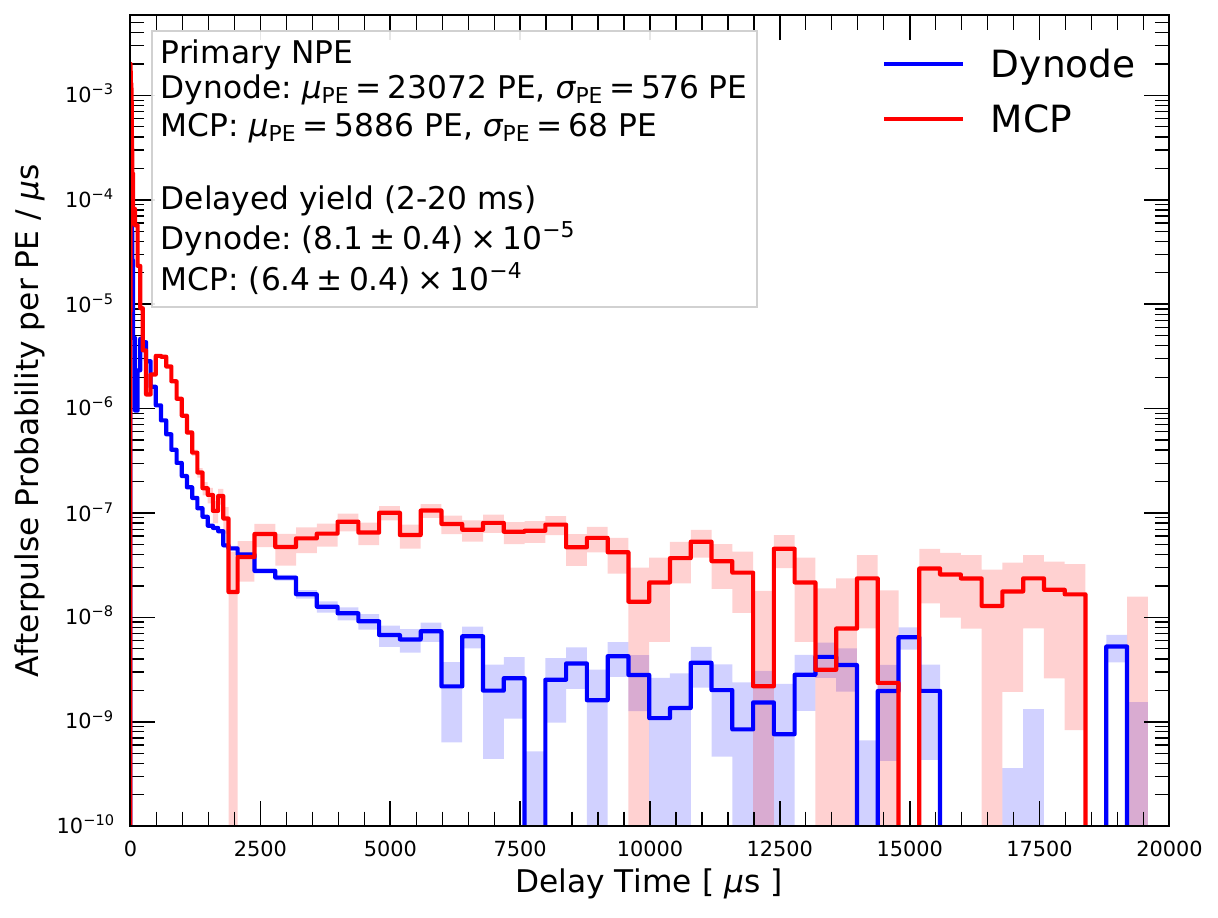}
\caption{
Afterpulse time profiles measured up to 20~ms for the dynode PMT and
the MCP-PMT using the sliding-window method, after subtraction of the pre-peak
dark-noise baseline. The distributions are normalized by the calibrated number
of photoelectrons in the primary pulse and by the bin width.
The inset lists the primary NPE, event-to-event standard deviation, and the integrated afterpulse
probabilities normalized to the primary NPE in the 2.0--20.0~ms window.
The uncertainties of the integrated probabilities include the mean-SPE calibration error;
curve bands contain profile and dark-noise statistics. The quoted $\sigma_{PE}$ values represent event-to-event fluctuations of the primary signal distribution.
}
\label{fig:time_profile_20ms}
\end{figure}

The resulting afterpulse time profiles are shown in
Figure~\ref{fig:time_profile_20ms}. The sliding-window readout extends the measurement to 20~ms while retaining
the 1~GS/s sampling rate within each recorded window.

The long-delayed afterpulse components at several hundred microseconds, consistent with
those observed in the 1.8~ms measurement, are still resolved with the
sliding-window method. This consistency supports the reliability of the extended
measurement. In the region later than 2~ms, the dynode PMT mainly shows the
falling tail of the broad component peaking around 260~$\mu$s, whereas the
MCP-PMT shows an additional millisecond-scale afterpulse component. After
normalization to the primary NPE, the integrated afterpulse probabilities in the
2.0--20.0~ms window are $(8.1\pm0.4)\times10^{-5}$ for the dynode PMT and
$(6.4\pm0.4)\times10^{-4}$ for the MCP-PMT, showing that these late
contributions remain very small. The uncertainties include the mean-SPE calibration error. For the MCP-PMT, the mean delay of this
component in the same window is 6.9~ms, with a standard deviation of 2.6~ms.
The extended measurement therefore quantifies a small, broadly distributed
afterpulse contribution in the millisecond-scale window.

\section{Discussion}
\label{sec:discuss}
\subsection{Large-Signal Response and Gain Recovery}
\label{sec:large_signal_gain_recovery}

The absolute normalization depends on the large-signal response of the PMT. Dedicated measurements of JUNO 20-inch PMTs have shown charge nonlinearity under intense pulsed illumination~\cite{Wu:2023linearity}. In this work, a split-fiber PD measurement directly monitors the incident light. The dynode PMT is consistent with a linear response over the analyzed range. For the MCP-PMT, the measured PD-current relation is used to convert each primary signal to the low-intensity linear-response scale. Residual uncertainty from this mapping is propagated to the horizontal error bars.

For the MCP-PMT, post-pulse gain suppression can instead bias the reconstructed afterpulse yield downward. The JUNO 20-inch MCP-PMTs use ALD-coated microchannels, which can exhibit temporary
gain suppression followed by recovery after large signals or high-rate
illumination~\cite{Li:2025mcp_saturation,Melikyan:2020load_capacity,Komarek:2021rate_capability}.
This may reduce the reconstructed afterpulse yield if some afterpulse hits fall
below the fixed reconstruction threshold. Therefore, large-signal effects mainly
affect the absolute yield normalization, while their impact on the measured time
components is expected to be smaller. The direct intensity scans were performed at low repetition rates, with a rate of 2 Hz for the MCP-PMT, to reduce inter-flash accumulation of gain-suppression effects.

\subsection{Higher-Order Afterpulse Contributions}

At high light intensity, an afterpulse with a sufficiently large avalanche may
itself generate further afterpulses. Such higher-order afterpulse contributions have
been considered in previous PMT studies~\cite{Haser:2013r7081_afterpulse}.
Unlike an uncorrelated dark-noise background, this contribution is tied to the
primary-induced afterpulse population and can therefore become more visible at
high light intensity. It may partly explain the deviations observed at high
primary light intensities and the residuals from the linear fits in
Figure~\ref{fig:late_ap_linearity},
although large-signal gain suppression and reconstruction nonlinearities may
also contribute. A quantitative treatment of this effect will be left to future
work.

\subsection{Physical Origins of the Long-Delayed Components}

The physical origins of the long-delayed components are not yet clear.
Conventional PMT afterpulses are often attributed to ion feedback,
where residual gas ions created by the primary avalanche drift back and release
secondary electrons~\cite{PBCoates_1973,Akchurin:2007ion_afterpulse,Ma:2011large_pmt_afterpulse}.
This mechanism has also been used to interpret short-delayed afterpulse groups
in the 20-inch HQE-MCP-PMT through flight-time arguments~\cite{WU2021165351}.
However, afterpulses on the hundreds-of-microseconds timescale and other long-delayed
backgrounds have been reported in different PMTs
\cite{POLESHCHUK2012362,TUDYKA201639,Bristow:2002sib}, and delayed electron
emission may also contribute to long-delayed components~\cite{Morozov:2023delayed_emission}.
The several-hundred-microsecond structures observed here may therefore arise
from ion feedback in weak-field or remote regions, delayed emission from PMT
surfaces, light feedback, or a combination of these processes.

Disentangling these possibilities requires dedicated measurements. High-voltage
scans can test ion time-of-flight scaling, while temperature scans may help
separate thermally assisted delayed emission from ion drift. Measurements with
different primary intensities, illumination positions, and wavelengths, combined
with joint time-amplitude-charge analysis, would provide stronger constraints on
the mechanism of these long-delayed components.

\section{Conclusion}
\label{sec:conclusion}

In this work, long-delayed afterpulse components in two 20-inch PMT models,
namely the HPK R12860 and NNVT GDB-6201, were measured using large primary
light signals and waveform reconstruction with dark-noise subtraction. The 1.8~ms DAQ-window measurement shows clear
afterpulse components beyond the conventional short-delay region. The dynode PMT
exhibits a broad component peaking at about 260~$\mu$s, followed by a slowly
decreasing tail, while the MCP-PMT shows a pronounced component around
90~$\mu$s and a weaker structure near 550~$\mu$s. These different time profiles
indicate that the long-delayed afterpulse response depends on the PMT
multiplication structure.

The yield dependence on the primary signal intensity was quantified by
integrating selected time windows from 10 to 1800~$\mu$s. The fitted
afterpulse slopes per primary PE are $(1.092\pm0.021)\times10^{-2}$ and
$(9.9\pm0.9)\times10^{-4}$ for the dynode PMT in the 10--100 and
100--1000~$\mu$s windows, respectively, and $(2.97\pm0.12)\times10^{-2}$,
$(5.8\pm0.3)\times10^{-3}$, and $(1.8\pm1.2)\times10^{-3}$ for the MCP-PMT in
the 10--50, 50--350, and 350--1800~$\mu$s windows, respectively. The increase with primary intensity supports the correlation of these components with the primary signal, although neither the dynode 10--100~$\mu$s nor the MCP-PMT 10--50~$\mu$s component is described by a single affine relation over the full measured range.

Using the sliding-window method, the measurement was further extended to 20~ms
while maintaining the 1~GS/s waveform sampling rate within each recorded window.
The several-hundred-microsecond components observed in the 1.8~ms
measurement are reproduced, supporting the reliability of the extended
measurement. In the 2.0--20.0~ms window, the afterpulse probabilities normalized
to the primary NPE are $(8.1\pm0.4)\times10^{-5}$ for the dynode PMT and
$(6.4\pm0.4)\times10^{-4}$ for the MCP-PMT, with statistical errors including the shared
mean-SPE calibration uncertainty. For the MCP-PMT, the
millisecond-scale component has a mean delay of 6.9~ms with a standard deviation
of 2.6~ms, showing that this small contribution is broadly distributed in time.

The absolute afterpulse yields are subject to large-signal systematic effects:
residual primary-scale uncertainty can affect the normalization, while post-pulse gain suppression and threshold effects can reduce the reconstructed afterpulse count. Higher-order afterpulse contributions may also lead to deviations from
linear light-intensity dependence at high primary signals. These effects will
require dedicated studies, but they are not expected to change the main timing
features reported here. The results demonstrate that sub-percent long-delayed
afterpulses extending from hundreds of microseconds to the millisecond scale
should be considered in PMT characterization and in the evaluation of correlated
backgrounds following high-intensity events in large scintillation detectors.

\acknowledgments

The authors thank Xilei Sun for suggesting the use of the CAEN DT5751 digitizer
for long-window waveform measurements, and Xuan Zhang for providing the
LabVIEW-based DT5751 readout software, which enabled millisecond-scale waveform
acquisition. The authors also thank Sen Qian and North Night Vision Technology Co., Ltd. for performing cross-check measurements of afterpulse components in the several-hundred-microsecond region and Guofu Cao for hardware support. AI-assisted tools were used solely for language editing and proofreading during manuscript preparation. All scientific content and results were independently verified by the authors.


\bibliographystyle{JHEP}
\bibliography{biblio.bib}

@manual{hamamatsu2006,
  title = {PHOTOMULTIPLIER TUBES: Basics and Applications},
  author = {{Hamamatsu Photonics K.K. Editorial Committee}},
  editor = {Hakamata, Toshikazu and Kume, Hidehiro and Okano, Kazuyoshi and Tomiyama, Kimiyuki and Kamiya, Akifumi and Yoshizawa, Yuji and Matsui, Hisayuki and Otsu, Ichiro and Taguchi, Takeshi and Kawai, Yoshihiko and Yamaguchi, Haruhisa and Suzuki, Kazumi and Suzuki, Seiji and Morita, Tetsuya},
  year = {2006},
  edition = {Third},
  organization = {Hamamatsu Photonics K.K., Electron Tube Division},
  address = {Japan},
  note = {Editing by Word Technical Writing, Inc. Comprehensive technical handbook on photomultiplier tubes (PMTs). Document code: TOTH9001E03. Available at: \url{https://psec.uchicago.edu/links/pmt_handbook_complete.pdf}},
  howpublished = {Technical Handbook}
}

@book{tsoulfanidis2010,
  title = {Measurement and Detection of Radiation},
  author = {Tsoulfanidis, Nicholas and Landsberger, Sheldon},
  year = {2010},
  edition = {Third},
  publisher = {CRC Press},
  address = {Boca Raton, FL},
  isbn = {978-1-4200-9157-1},
  note = {A comprehensive introduction to radiation detection and measurement for nuclear science and engineering. Covers PMTs in Chapter 8.}
}

@article{Liu:2025evh,
    author = "Liu, Caimei and others",
    title = "{{JUNO} 20-inch PMT and electronics system characterization using large pulses of PMT dark counts at the Pan-Asia testing platform}",
    journal = {JINST},
    volume = "20",
    number = "12",
    pages = "P12013",
    year = "2025",
    doi = "10.1088/1748-0221/20/12/P12013",
    eprint = "2506.21179",
    archivePrefix = "arXiv",
    primaryClass = "physics.ins-det"
}

@article{Li:2025mcp_saturation,
    author = "Li, Kuinian and others",
    title = "{Experimental investigation of saturation recovery behavior in MCP-PMT}",
    journal = {Nucl. Instrum. Meth. A},
    volume = "1074",
    pages = "170323",
    year = "2025",
    doi = "10.1016/j.nima.2025.170323"
}

@article{Wu:2023linearity,
    author = "Wu, Diru and Luo, Fengjiao and Wang, Zhimin and Li, Min and Xu, Jilei and He, Miao and Yang, Changgen and Heng, Yuekun",
    title = "{Study on the linearity of 20'' dynode and MCP PMTs}",
    journal = {JINST},
    volume = "18",
    number = "05",
    pages = "P05033",
    year = "2023",
    doi = "10.1088/1748-0221/18/05/P05033",
    eprint = "2212.11514",
    archivePrefix = "arXiv",
    primaryClass = "physics.ins-det"
}

@article{Melikyan:2020load_capacity,
    author = "Melikyan, Y. and others",
    title = "{Load capacity and recovery behaviour of ALD-coated MCP-PMTs}",
    journal = {Nucl. Instrum. Meth. A},
    volume = "949",
    pages = "162854",
    year = "2020",
    doi = "10.1016/j.nima.2019.162854"
}

@article{Komarek:2021rate_capability,
    author = "Komarek, T. and others",
    title = "{Timing resolution and rate capability of Photonis miniPlanacon XPM85212/A1-S MCP-PMT}",
    journal = {Nucl. Instrum. Meth. A},
    volume = "985",
    pages = "164705",
    year = "2021",
    doi = "10.1016/j.nima.2020.164705"
}

@article{Haser:2013r7081_afterpulse,
doi = {10.1088/1748-0221/8/04/P04029},
url = {https://doi.org/10.1088/1748-0221/8/04/P04029},
year = {2013},
month = {apr},
publisher = {},
volume = {8},
number = {04},
pages = {P04029},
author = {J Haser and F Kaether and C Langbrandtner and M Lindner and S Lucht and S Roth and M Schumann and A Stahl and A Stüken and C Wiebusch},
title = {Afterpulse measurements of {R7081} photomultipliers for the {Double Chooz} experiment},
journal = {JINST}
}

@article{Ma:2011large_pmt_afterpulse,
title = {Time and amplitude of afterpulse measured with a large size photomultiplier tube},
journal = {Nucl. Instrum. Meth. A},
volume = {629},
number = {1},
pages = {93--100},
year = {2011},
issn = {0168-9002},
doi = {10.1016/j.nima.2010.11.095},
url = {https://www.sciencedirect.com/science/article/pii/S0168900210026306},
author = {K.J. Ma and W.G. Kang and J.K. Ahn and S. Choi and Y. Choi and M.J. Hwang and J.S. Jang and E.J. Jeon and K.K. Joo and H.S. Kim and J.Y. Kim and S.B. Kim and S.H. Kim and W. Kim and Y.D. Kim and J. Lee and I.T. Lim and Y.D. Oh and M.Y. Pac and C.W. Park and I.G. Park and K.S. Park and S.S. Stepanyan and I. Yu}
}

@article{Bristow:2002sib,
    author  = {Bristow, Michael P.},
    title   = {Suppression of afterpulsing in photomultipliers
               by gating the photocathode},
    journal = {Appl. Opt.},
    volume  = {41},
    number  = {24},
    pages   = {4975--4987},
    year    = {2002},
    doi     = {10.1364/AO.41.004975}
}

@article{Akchurin:2007ion_afterpulse,
    author = "Akchurin, Nural and Kim, Heejong",
    title = "{A study on ion initiated photomultiplier afterpulses}",
    journal = {Nucl. Instrum. Meth. A},
    volume = "574",
    number = "1",
    pages = "121--126",
    year = "2007",
    doi = "10.1016/j.nima.2007.01.093"
}

@article{Genster_2020,
doi = {10.1088/1742-6596/1342/1/012116},
url = {https://doi.org/10.1088/1742-6596/1342/1/012116},
year = {2020},
month = {jan},
publisher = {IOP Publishing},
volume = {1342},
number = {1},
pages = {012116},
author = {Genster, Christoph},
title = {Studies on Muon Veto in the {JUNO} Liquid Scintillator Neutrino Detector},
journal = {J. Phys. Conf. Ser.}
}

@article{Morozov:2023delayed_emission,
    author = "Morozov, V. A. and Morozova, N. V. and Budzynski, P.",
    title = "{Delayed electron emission in photomultiplier tubes}",
    journal = {Nucl. Instrum. Meth. A},
    volume = "1053",
    pages = "168323",
    year = "2023",
    doi = "10.1016/j.nima.2023.168323"
}

@article{RENKER2009207,
title = {New developments on photosensors for particle physics},
journal = {Nucl. Instrum. Meth. A},
volume = {598},
number = {1},
pages = {207--212},
year = {2009},
note = {Instrumentation for Collding Beam Physics},
issn = {0168-9002},
doi = {10.1016/j.nima.2008.08.023},
url = {https://www.sciencedirect.com/science/article/pii/S016890020801228X},
author = {D. Renker}
}

@article{kubetsky1937,
author = {Kubetsky, L. A.},
title = {Multiple Amplifier},
journal = {Proc. IRE},
volume = {25},
number = {4},
pages = {421},
year = {1937},
doi = {10.1109/JRPROC.1937.229045}
}

@article{zworykin1936,
author = {Zworykin, V. K. and Morton, G. A. and Malter, L.},
title = {The Secondary Emission Multiplier—A New Electronic Device},
journal = {Proc. IRE},
volume = {24},
number = {3},
pages = {351},
year = {1936},
doi = {10.1109/JRPROC.1936.226435}
}

@article{anger1958,
  author = {Anger, H. O.},
  title = {Scintillation Camera},
  journal = {Rev. Sci. Instrum.},
  volume = {29},
  number = {1},
  pages = {27--33},
  year = {1958},
  doi = {10.1063/1.1715998}
}

@book{knoll1999,
  author = {Knoll, G. F.},
  title = {Radiation Detection and Measurement},
  edition = {3},
  publisher = {John Wiley \& Sons, Inc.},
  year = {1999},
  isbn = {978-0471495456}
}

@article{Alimonti:2009aa,
  author = {Alimonti, G. and others},
  title = {The {Borexino} detector at the {Laboratori Nazionali del Gran Sasso}},
  journal = {Nucl. Instrum. Meth. A},
  volume = {600},
  year = {2009},
  pages = {568--593},
  doi = {10.1016/j.nima.2008.11.056},
  eprint = {0806.2400},
  archivePrefix = {arXiv},
  primaryClass = {physics.ins-det}
}

@article{Eguchi:2002nm,
  author = {Eguchi, K. and others},
  title = {First results from {KamLAND}: Evidence for reactor antineutrino disappearance},
  journal = {Phys. Rev. Lett.},
  volume = {90},
  year = {2003},
  pages = {021802},
  doi = {10.1103/PhysRevLett.90.021802},
  eprint = {hep-ex/0212021},
  archivePrefix = {arXiv},
  primaryClass = {hep-ex}
}

@article{An:2015kca,
  author = {{JUNO collaboration}},
  title = {Neutrino physics with {JUNO}},
  journal = {J. Phys. G},
  volume = {43},
  year = {2016},
  pages = {030401},
  doi = {10.1088/0954-3899/43/3/030401},
  eprint = {1507.05613},
  archivePrefix = {arXiv},
  primaryClass = {physics.ins-det}
}

@article{P_B_Coates_1973,
doi = {10.1088/0022-3727/6/16/306},
url = {https://dx.doi.org/10.1088/0022-3727/6/16/306},
year = {1973},
month = {oct},
publisher = {},
volume = {6},
number = {16},
pages = {1862},
author = {P B Coates},
title = {A theory of afterpulse formation in photomultipliers and the prepulse height distribution},
journal = {J. Phys. D: Appl. Phys.},
}

@article{PhysRev.84.1248,
  title = {Satellite Pulses from Photomultipliers},
  author = {Godfrey, T. N. K. and Harrison, F. B. and Keuffel, J. W.},
  journal = {Phys. Rev.},
  volume = {84},
  issue = {6},
  pages = {1248--1249},
  numpages = {0},
  year = {1951},
  month = {Dec},
  publisher = {American Physical Society},
  doi = {10.1103/PhysRev.84.1248},
  url = {https://link.aps.org/doi/10.1103/PhysRev.84.1248}
}

@article{Zhao_2016,
doi = {10.1088/1748-0221/11/05/T05002},
url = {https://dx.doi.org/10.1088/1748-0221/11/05/T05002},
year = {2016},
month = {may},
publisher = {},
volume = {11},
number = {05},
pages = {T05002},
author = {Zhao, X. and Tang, Z. and Li, C. and Chen, H. and Zhang, Y. and Li, X. and Shao, M. and Sun, Y. and Zha, W. and Zhou, Y.},
title = {Afterpulse measurement for 8-inch candidate {PMTs} for {LHAASO}},
journal = {JINST}
}

@article{Zhao:2022gks,
    author = "Zhao, Rong and others",
    title = "{Afterpulse measurement of {JUNO} 20-inch PMTs}",
    eprint = "2207.04995",
    archivePrefix = "arXiv",
    primaryClass = "physics.ins-det",
    doi = "10.1007/s41365-022-01162-3",
    journal = {Nucl. Sci. Tech.},
    volume = "34",
    number = "1",
    pages = "12",
    year = "2023"
}

@article{WU2021165351,
title = {Study of after-pulses in the 20-inch {HQE-MCP-PMT} for the {JUNO} experiment},
journal = {Nucl. Instrum. Meth. A},
volume = {1003},
pages = {165351},
year = {2021},
issn = {0168-9002},
doi = {10.1016/j.nima.2021.165351},
url = {https://www.sciencedirect.com/science/article/pii/S0168900221003351},
author = {Qi Wu and Sen Qian and Lishuang Ma and Jingkai Xia and Bayarto K. Lubsandorzhiev and Zhigang Wang and Yao Zhu and Haitao Li and Nikita Ushakov and Andrei Sidorenkov and Qianyu Hu and Jianning Sun and Shuguang Si}
}

@article{TUDYKA201639,
title = {A study on photomultiplier afterpulses in {TL/OSL} readers},
journal = {Radiat. Meas.},
volume = {86},
pages = {39--48},
year = {2016},
issn = {1350-4487},
doi = {10.1016/j.radmeas.2016.01.004},
url = {https://www.sciencedirect.com/science/article/pii/S135044871630004X},
author = {Konrad Tudyka and Andrzej Bluszcz}
}

@article{POLESHCHUK2012362,
title = {An observation of a new class of afterpulses with delay time in the range of 70–200 $\mu$s in classical vacuum photomultipliers},
journal = {Nucl. Instrum. Meth. A},
volume = {695},
pages = {362--364},
year = {2012},
note = {New Developments in Photodetection NDIP11},
issn = {0168-9002},
doi = {10.1016/j.nima.2011.11.030},
url = {https://www.sciencedirect.com/science/article/pii/S0168900211020729},
author = {R.V. Poleshchuk and B.K. Lubsandorzhiev and R.V. Vasiliev}
}

@article{abusleme_mass_2022,
    author = "Abusleme, Angel and others",
    collaboration = "JUNO",
    title = "{Mass testing and characterization of 20-inch PMTs for {JUNO}}",
    journal = {Eur. Phys. J. C},
    volume = "82",
    number = "12",
    pages = "1168",
    year = "2022",
    doi = "10.1140/epjc/s10052-022-11002-8"
}

@article{PBCoates_1973,
doi = {10.1088/0022-3727/6/10/301},
url = {https://doi.org/10.1088/0022-3727/6/10/301},
year = {1973},
month = {jun},
publisher = {},
volume = {6},
number = {10},
pages = {1159},
author = {P B Coates},
title = {The origins of afterpulses in photomultipliers},
journal = {J. Phys. D: Appl. Phys.}
}

@article{IceCube:2025paj,
    author = "Dutta, Kaustav and others",
    collaboration = "IceCube",
    title = "{Very Late Afterpulses and Search for the Neutron Echo in IceCube}",
    eprint = "2507.07042",
    archivePrefix = "arXiv",
    primaryClass = "astro-ph.HE",
    reportNumber = "PoS-ICRC2025-1030",
    doi = "10.22323/1.501.1030",
    journal = {PoS},
    volume = "ICRC2025",
    pages = "1030",
    year = "2025"
}

@article{IceCube:2013cd,
    author = "Aartsen, M. G. and others",
    collaboration = "IceCube",
    title = "{Evidence for High-Energy Extraterrestrial Neutrinos at the IceCube Detector}",
    eprint = "1311.5238",
    archivePrefix = "arXiv",
    primaryClass = "astro-ph.HE",
    journal = {Science},
    volume = "342",
    pages = "1242856",
    doi = "10.1126/science.1242856",
    year = "2013"
}

@article{LHAASO:2021gok,
    author = "Cao, Zhen and others",
    collaboration = "LHAASO",
    title = "{Ultrahigh-energy photons up to 1.4 petaelectronvolts from 12 \ensuremath{\gamma}-ray Galactic sources}",
    journal = {Nature},
    volume = "594",
    number = "7861",
    pages = "33--36",
    doi = "10.1038/s41586-021-03498-z",
    year = "2021"
}

@article{KM3NeT:2025aa,
    author = "Aiello, S. and others",
    collaboration = "KM3NeT",
    title = "{Observation of an ultra-high-energy cosmic neutrino with KM3NeT}",
    journal = {Nature},
    volume = "638",
    number = "8056",
    pages = "376--382",
    doi = "10.1038/s41586-024-08543-1",
    year = "2025"
}

@article{juno_ppnp,
    author = {{JUNO collaboration}},
    collaboration = "JUNO",
    title = "{{JUNO} physics and detector}",
    eprint = "2104.02565",
    archivePrefix = "arXiv",
    primaryClass = "hep-ex",
    doi = "10.1016/j.ppnp.2021.103927",
    journal = {Prog. Part. Nucl. Phys.},
    volume = "123",
    pages = "103927",
    year = "2022"
}

@article{juno_reactor_nmo,
    author = {{JUNO collaboration}},
    collaboration = "JUNO",
    title = "{Potential to identify neutrino mass ordering with reactor antineutrinos at {JUNO}}",
    eprint = "2405.18008",
    archivePrefix = "arXiv",
    primaryClass = "hep-ex",
    doi = "10.1088/1674-1137/ad7f3e",
    journal = {Chin. Phys. C},
    volume = "49",
    number = "3",
    pages = "033104",
    year = "2025"
}

@article{juno_initial_performance_2026,
doi = {10.1088/1674-1137/ae3dc1},
url = {https://doi.org/10.1088/1674-1137/ae3dc1},
year = {2026},
month = {apr},
publisher = {Chinese Physical Society and the Institute of High Energy Physics of the Chinese Academy of Sciences and the Institute of Modern Physics of the Chinese Academy of Sciences and IOP Publishing Ltd
				},
volume = {50},
number = {4},
pages = {043001},
author = {{JUNO collaboration}},
title = {Initial performance results of the {JUNO} detector},
journal = {Chin. Phys. C}
}

@article{JUNO:2025first_osc,
  author  = {{JUNO collaboration}},
  title   = {Measurement of reactor neutrino oscillation with the first {JUNO} data},
  journal = {Nature},
  year    = {2026},
  date    = {2026-06-01},
  volume  = {654},
  number  = {8118},
  pages   = {343--348},
  doi     = {10.1038/s41586-026-10538-z},
  url     = {https://doi.org/10.1038/s41586-026-10538-z},
  issn    = {1476-4687}
}

@article{Luo_2025,
doi = {10.1088/1748-0221/20/07/P07024},
url = {https://doi.org/10.1088/1748-0221/20/07/P07024},
year = {2025},
month = {jul},
publisher = {IOP Publishing},
volume = {20},
number = {07},
pages = {P07024},
author = {Luo, Fengjiao and Wang, Zhimin and Yang, Anbo and Heng, Yuekun and Qin, Zhonghua and Xu, Meihang and Qian, Sen and Liu, Shulin and Wang, Yifang and Wang, Wei and Olshevskiy, Alexander and Huang, Guorui and Jin, Zhen and Ren, Ling and Wang, Xingchao and Si, Shuguang and Sun, Jianning},
title = {Design \& optimization of the {HV} divider for {JUNO} 20-inch {PMT}},
journal = {JINST}
}

@article{Saldanha:2017spe,
    author = {Saldanha, R. and Grandi, L. and Guardincerri, Y. and Wester, T.},
    title = {{Model Independent Approach to the Single Photoelectron Calibration of Photomultiplier Tubes}},
    journal = {Nucl. Instrum. Meth. A},
    volume = {863},
    pages = {35--46},
    year = {2017},
    doi = {10.1016/j.nima.2017.02.086},
    eprint = {1602.03150},
    archivePrefix = {arXiv},
    primaryClass = {physics.ins-det}
}
\end{document}